\documentclass[twocolumn,pre,superscriptaddress]{revtex4}

\usepackage{dcolumn}
\usepackage{amsmath}

\usepackage{graphicx}
\usepackage{epstopdf}

\newcommand{\e}{\mathrm{e}}

\newcommand{\av}[1]{\left\langle#1\right\rangle}
\newcommand{\etal}{{\it{}et~al.}}
\newcommand{\defn}{\textit}
\newcommand{\Ord}{\mathrm{O}}

\renewcommand{\vec}{\mathbf}

\begin{document}
\title{Identification of core-periphery structure in networks}

\author{Xiao Zhang}
\affiliation{Department of Physics, University of Michigan, Ann Arbor, MI
  48109}
\author{Travis Martin}
\affiliation{Department of Electrical Engineering and Computer Science,
  University of Michigan, Ann Arbor, MI 48109}
\author{M. E. J. Newman}
\affiliation{Department of Physics, University of Michigan, Ann Arbor, MI
48109}
\affiliation{Center for the Study of Complex Systems, University of
Michigan, Ann Arbor, MI 48109}

\begin{abstract}
  Many networks can be usefully decomposed into a dense core plus an
  outlying, loosely-connected periphery.  Here we propose an algorithm for
  performing such a decomposition on empirical network data using methods
  of statistical inference.  Our method fits a generative model of
  core--periphery structure to observed data using a combination of an
  expectation--maximization algorithm for calculating the parameters of the
  model and a belief propagation algorithm for calculating the
  decomposition itself.  We find the method to be efficient, scaling easily
  to networks with a million or more nodes and we test it on a range of
  networks, including real-world examples as well as computer-generated
  benchmarks, for which it successfully identifies known core--periphery
  structure with low error rate.  We also demonstrate that the method is
  immune from the detectability transition observed in the related
  community detection problem, which prevents the detection of community
  structure when that structure is too weak.  There is no such transition
  for core--periphery structure, which is detectable, albeit with some
  statistical error, no matter how weak it is.
\end{abstract}

\maketitle

\section{Introduction}
\label{sec:intro}
Much of the recent work on the structure of networked systems, such as
social and technological networks, has focused on measurements of local
structure, such as vertex degrees, clustering coefficients, correlations,
and so forth~\cite{Boccaletti06,Newman10}.  Increasingly, however,
researchers have investigated medium- and large-scale structure as well.
The lion's share of the attention has gone to the study of so-called
community structure~\cite{Fortunato10}, the archetypal example of
large-scale network structure, in which the nodes of a network are divided
into tightly knit groups or communities that often reflect aspects of
network function.  But other structure types can be important as well and
recent research has also looked at overlapping or fuzzy
communities~\cite{PDFV05,ABFX08,ABL10}, hierarchical
structure~\cite{RB03,CMN08}, and ranking~\cite{BN13}, among others.

In this paper we focus on another, distinct type of large-scale structure,
\defn{core--periphery structure}.  Many networks are observed to divide
into a densely interconnected core surrounded by a sparser halo or
periphery.  Already in the 1990s sociologists observed such structure in
social networks~\cite{BE99} and more recently a number of researchers have
made quantitative studies of core--periphery structure in a range of
different types of networks~\cite{Holme05b,RPFM14}.  The identification of
core--periphery structure has a number of potential uses.  Core nodes in a
network might play a different role from peripheral ones~\cite{GA05} and
the ability to distinguish core from periphery might thus give us a new
handle on function in networked systems.  Distinguishing between core and
periphery might lead to more informative visualizations of networks or find
a role in graph layout algorithms similar to that played today by community
structure.  And core nodes, for instance in social networks, might be more
influential or powerful than peripheral ones, so the ability to discern the
difference could shed light on social or other organization.

There have also been studies of two other types of structure that are
reminiscent of, though different in important ways from, core--periphery
structure: \defn{``rich club'' structure}~\cite{CFSV06,ZM04} and
\defn{degree assortativity}~\cite{PVV01,Newman02f}.  A rich-club is a group
of high-degree nodes in a network (i.e.,~nodes with many connections to
others) that preferentially connect to one another.  Such a club is a
special case of the core in a core--periphery structure, but the concept of
a core is more general, encompassing cases (as we will see) in which
low-degree nodes can also belong to the core.  The rich-club phenomenon
also makes no statement about connectivity patterns in the remainder of the
network, where as core--periphery structure does.

Assortative mixing is the tendency of nodes in a network to connect to
others that are similar to themselves in some way, and degree-assortative
mixing is the tendency to connect to others with similar degree---high to
high, and low to low.  This produces a core in the network of connected
high-degree vertices, similar to the rich-club, but low-degree vertices
also preferentially connect to one another and prefer not to connect to the
core, which is the opposite of core--periphery structure as commonly
understood, in which periphery vertices are more likely to connect to the
core than they are to one another.

A number of suggestions have been made about how, given the complete
pattern of connections in a network, one could detect core--periphery
structure in that pattern.  All of them take the same basic approach of
defining an objective function that measures the strength or quality of a
candidate division into core and periphery and then maximizes (usually only
approximately) over divisions to find the best one.  In early work,
Borgatti and Everett~\cite{BE99} proposed a quality function based on
comparing the network to an ideal core--periphery model in which nodes are
connected to each other if and only if they are members of the core.
Rombach~\etal~\cite{RPFM14} built on the same idea, but using a more
flexible model.  Holme~\cite{Holme05b} took a contrasting approach
reminiscent of the clustering coefficient used to quantify transitivity in
networks.

In this paper we propose a different, statistically principled method of
detecting core--periphery structure using a maximum-likelihood fit to a
generative network model.  The method is conceptually similar to
recently-popular first-principles methods for community
detection~\cite{BC09,DKMZ11a} and in fact uses the same underlying network
model, the stochastic block model, although with a different choice of
parameters appropriate to core--periphery rather than community structure.
Among other results we demonstrate that the method is able consistently to
detect planted core--periphery structure in computer-generated test
networks, and that, by contrast with the community detection problem, there
is no minimum amount of structure that can be detected.  Any
core--periphery structure, no matter how weak, is in principle detectable.

\section{The stochastic block model}
The stochastic block model is a well established and widely used model for
community structure in networks.  It is a generative model, meaning its
original purpose is to create artificial networks that contain community
structure.  It is also commonly used, however, for community detection by
fitting the model to observed network data.  The parameters of the fit tell
us the best division of the network into communities.

The model is defined as follows.  We take $n$ nodes, initially without any
edges connecting them, and divide them into some number of groups.  We will
consider the simplest case where there are just two groups (which will
represent the core and periphery).  Each vertex is assigned randomly to
group~1 with probability~$\gamma_1$ or group~2 with probability~$\gamma_2 =
1-\gamma_1$.  Then between every vertex pair we place an undirected edge
independently at random with probability~$p_{rs}$, or not with
probability~$1-p_{rs}$, where $r$ and $s$ are the groups to which the two
vertices belong.  Thus the probability of connection of any two vertices
depends solely on their group membership.  The probabilities~$p_{rs}$ form
a matrix, sometimes called the \defn{mixing matrix} or \defn{affinity
  matrix}, which is a $2\times2$ matrix in our two-group example.  Since
the edges in the network are undirected it follows that the mixing matrix
is symmetric, $p_{12}=p_{21}$, leaving three independent probabilities that
we can choose, $p_{11}$, $p_{12}$, and~$p_{22}$.

In the most commonly studied case the probabilities for connection within
groups are chosen to be larger than the probabilities between groups
$p_{11}>p_{12}<p_{22}$.  This gives traditional community structure, also
called assortative mixing, with denser connections within groups than
between them.  A contrasting possibility is the disassortative choice
$p_{11}<p_{12}>p_{22}$, where edges are more probable between groups than
within them.  This choice, and the structure it describes, has received a
modest amount of attention in the literature~\cite{Newman03c,YZRM11}.

There is, however, a third possibility that has rarely been studied, in
which $p_{11}>p_{12}>p_{22}$.  This is the situation we refer to as
core--periphery structure.  Since the group labels are arbitrary we can,
without loss of generality, assume $p_{11}$ to be the largest of the three
probabilities, so group~1 is the core.  Connections are most probable
within the core, least probable within the periphery, and of intermediate
probability between core and periphery.  Note that this means that
periphery vertices are more likely to be connected to core vertices than to
each other, a characteristic feature of core--periphery structure that
distinguishes it from either assortative or disassortative mixing.

As we have said, the stochastic block model can be used to detect structure
in network data by fits of the data to the model.  For instance, the
assortative version of the model can be used to fit and hence detect
community structure in networks~\cite{BC09,DKMZ11a}.  As shown
in~\cite{KN11a}, however, it often performs poorly at this task in
real-world situations because real-world networks tend to have broad degree
distributions that dominate the large-scale structure and the fit tends to
pick out this gross effect rather than the more subtle underlying community
structure---typically the fit just ends up dividing the network into groups
of higher- and lower-degree vertices rather than traditional communities.
A more nuanced view has been given by Decelle~\etal~\cite{DKMZ11b}, who
show that in fact both the degree-based division and the community division
are good fits to the model---local maxima of the likelihood in the
language introduced below---but the degree-based one is better.

But when we turn to core--periphery structure this bug becomes a feature.
In networks with core--periphery structure the vertices in the core
typically do have higher degree than those in the periphery, so a method
that recognizes this fact is doing the right thing.  Indeed, as we show in
Section~\ref{sec:examples}, one can in certain cases do a reasonable job of
detecting core--periphery structure just by separating vertices into two
groups according to their degrees.  On the other hand, one can do better
still using the stochastic block model.

\section{Fitting to empirical data}
We propose to detect core--periphery structure in networks by finding the
parameters of the stochastic block model that best fit the model to a given
observed network.  This we do by the method of maximum likelihood,
implemented using an expectation--maximization or EM
algorithm~\cite{DLR77}.  The use of EM algorithms for network model fitting
is well established~\cite{NS01,NL07}, but it is worth briefly running
through the derivation for our particular model, which goes as follows.

\subsection{The EM algorithm}
\label{sec:em}
Given a network, the question we ask is, if this network were generated by
the stochastic block model, what is our best guess at the values of the
parameters of that model?  To answer this question, let $A_{ij}$ be an
element of the adjacency matrix~$A$ of the network having value one if
there is an edge between vertices $i$ and~$j$ and zero otherwise, and let
$g_i$ be the group that vertex~$i$ belongs to.  Then the probability, or
likelihood, that the network was generated by the model is
\begin{align}
P(A|p,\gamma) &= \sum_g P(A|p,\gamma,g) P(g|\gamma) \nonumber\\
  &= \sum_g \prod_{i<j} p_{g_ig_j}^{A_{ij}} (1 - p_{g_ig_j})^{1-A_{ij}}
            \prod_i \gamma_{g_i},
\end{align}
where $\sum_g$ indicates a sum over all assignments of the vertices to
groups.

To determine the most likely values of the parameters~$p_{rs}$
and~$\gamma_r$, we maximize this likelihood with respect to them.  In fact
it is technically simpler to maximize the logarithm of the likelihood:
\begin{equation}
\log P(A|p,\gamma) = \log \sum_g \prod_{i<j} p_{g_ig_j}^{A_{ij}}
        (1 - p_{g_ig_j})^{1-A_{ij}} \prod_i \gamma_{g_i},
\label{eq:loglikelihood}
\end{equation}
which is equivalent since the logarithm is a monotone increasing function.
Direct maximization is still quite difficult, however.  Simply
differentiating to find the maximum leads to a complex set of implicit
equations that have no easy solution.

A better approach, and the one taken in the EM algorithm, involves the
application of Jensen's inequality, which says that for any set of
positive-definite quantities~$x_i$
\begin{equation}
\log \sum_i x_i \ge \sum_i q_i \log {x_i\over q_i},
\end{equation}
where $q_i$ is any probability distribution satisfying the normalization
condition $\sum_i q_i = 1$.  One can easily verify that the exact equality
is achieved by choosing
\begin{equation}
q_i = x_i/\sum_i x_i.
\label{eq:equality}
\end{equation}

\begin{widetext}
  For any properly normalized probability distribution~$q(g)$ over the
  group assignments~$g$, Jensen's inequality applied to
  Eq.~\eqref{eq:loglikelihood} gives
\begin{align}
\log P(A|p,\gamma) &\ge \sum_g q(g) \log \biggl[ {1\over q(g)}
     \prod_{i<j} p_{g_ig_j}^{A_{ij}} (1 - p_{g_ig_j})^{1-A_{ij}}
     \prod_i \gamma_{g_i} \biggr] \nonumber\\
  &= \sum_g q(g) \biggl[ \sum_{i<j} \bigl[ A_{ij} \log p_{g_ig_j} +
     (1-A_{ij}) \log(1-p_{g_ig_j}) \bigr] + \sum_i \log \gamma_{g_i}
     - q(g) \log q(g) \biggr] \nonumber\\
  &= \frac12 \sum_{ij} \sum_{rs} \bigl[ A_{ij}
      q_{rs}^{ij} \log p_{rs} + (1-A_{ij}) q_{rs}^{ij} \log(1-p_{rs})
      \bigr] + \sum_{ir} q_r^i \log \gamma_r - \sum_g q(g) \log q(g),
\label{eq:ineq}
\end{align}
\end{widetext}
where $q_r^i$ is the so-called \defn{marginal probability} within the
chosen distribution~$q(g)$ that vertex~$i$ belongs to group~$r$:
\begin{equation}
q_r^i = \sum_g q(g) \delta_{g_i,r},
\label{eq:onepoint}
\end{equation}
and $q_{rs}^{ij}$ is the joint or two-vertex marginal probability that
vertex~$i$ belongs to group~$r$ and vertex~$j$ simultaneously belongs to
group~$s$:
\begin{equation}
q_{rs}^{ij} = \sum_g q(g) \delta_{g_i,r} \delta_{g_j,s},
\label{eq:twopoint}
\end{equation}
with $\delta_{ij}$ being the Kronecker delta.

Following Eq.~\eqref{eq:equality}, the exact equality in~\eqref{eq:ineq} is
achieved when
\begin{equation}
q(g) = {\prod_{i<j} p_{g_ig_j}^{A_{ij}} (1 - p_{g_ig_j})^{1-A_{ij}}
        \prod_i \gamma_{g_i}\over
        \sum_g \prod_{i<j} p_{g_ig_j}^{A_{ij}} (1 - p_{g_ig_j})^{1-A_{ij}}
        \prod_i \gamma_{g_i}}.
\label{eq:estep}
\end{equation}
Thus calculating the maximum of the left-hand side of~\eqref{eq:ineq} with
respect to the parameters~$p,\gamma$ is equivalent to first maximizing the
right-hand side with respect to~$q(g)$ (by choosing the value above) so as
to make the two sides equal, and then maximizing the result with respect to
the parameters.  In this way we turn our original problem of maximizing
over the parameters into a double maximization of the right-hand side
expression over the parameters and the distribution~$q(g)$.  At first
glance, this would seem to make the problem more difficult, but numerically
it is in fact easier, since it splits a challenging maximization into two
separate and relatively elementary operations.  The maximization with
respect to the parameters is achieved by straightforward differentiation
of~\eqref{eq:ineq} with the constraint that $\sum_r \gamma_r = 1$.  Note
that the final term on the right-hand side does not depend on the
parameters and hence vanishes upon differentiation, and we arrive at the
following expressions for the parameters:
\begin{align}
\label{eq:prs0}
p_{rs}   &= {\sum_{ij} A_{ij} q_{rs}^{ij}\over \sum_{ij} q_{rs}^{ij}}, \\
\label{eq:gamma}
\gamma_r &= {1\over n} \sum_i q_r^i,
\end{align}
where $n$ is the total number of vertices as previously.  The simultaneous
solution of Eqs.~\eqref{eq:estep} to~\eqref{eq:gamma} now gives us the
optimal values of the parameters.

The EM algorithm solves these equations by numerical iteration.  Given an
initial guess at the parameters~$p$ and~$\gamma$ we can calculate the
probability distribution~$q(g)$ from Eq.~\eqref{eq:estep} and from it the
one- and two-vertex marginal probabilities, Eqs.~\eqref{eq:onepoint}
and~\eqref{eq:twopoint}.  And from these we can calculate a new estimate of
$p$ and $\gamma$ from Eqs.~\eqref{eq:prs0} and~\eqref{eq:gamma}.  It can be
proved that upon iteration this process will always converge to a local
maximum of the log-likelihood.  It may not be the global maximum, however,
so commonly one performs the entire calculation several times with
different starting conditions, choosing from among the solutions so
obtained the one with the highest likelihood.

Equation~\eqref{eq:prs0} can be simplified a little further by using
Eq.~\eqref{eq:twopoint} to rewrite the denominator thus:
\begin{equation}
\sum_{ij} q_{rs}^{ij}
  = \sum_g q(g) \sum_i \delta_{g_i,r} \sum_j \delta_{g_j,s}
  = \av{n_r n_s},
\end{equation}
where $\av{\ldots}$ indicates an average within the probability
distribution~$q(g)$ and $n_r=\sum_i \delta_{g_i,r}$ is the number of
vertices in group~$r$.  In the limit of large network size the number of
vertices in a group becomes narrowly peaked and we can replace $\av{n_r
  n_s}$ by $\av{n_r}\av{n_s}$ with
\begin{equation}
\av{n_r} = \sum_g q(g) \sum_i \delta_{g_i,r} = \sum_i q_r^i,
\end{equation}
where we have used Eq.~\eqref{eq:onepoint}.  Then
\begin{equation}
p_{rs} = {\sum_{ij} A_{ij} q_{rs}^{ij}\over\sum_i q_r^i \sum_j q_s^j}.
\label{eq:prs}
\end{equation}
This expression has the advantage of requiring only a sum over edges in the
numerator (since one need sum only those terms for which $A_{ij}=1$) and
single sums over vertices in the denominator, not the double sum in the
denominator of~\eqref{eq:prs0}.  This makes evaluation of~$p_{rs}$
significantly faster for large networks.  (Note, however, that despite
appearances, Eq.~\eqref{eq:prs} does not assume that
$q_{rs}^{ij}=q_r^iq_s^j$, which would certainly not be correct in general.
Only the sum over all vertex pairs factorizes, not the individual terms.)

The final result of the EM algorithm gives us not only the values of the
parameters, but also the marginal probabilities~$q_r^i$ for vertices to
belong to each group.  In fact, it is normally this latter quantity that we
are really interested in.  In the community structure context it gives the
probability that vertex~$i$ belongs to community~$r$.  In the
core--periphery case, it gives the probability that the vertex belongs to
either the core (group~1) or the periphery (group~2).  Typically, the last
step in the calculation is to assign each vertex to the group for which it
has highest probability of membership, producing the final division of the
network into core and periphery.

\subsection{Belief propagation} 
\label{sec:bp}
The EM algorithm is an elegant approach but it has its shortcomings.
Principal among them is the difficulty of performing the sum over group
assignments~$g$ in the denominator of Eq.~\eqref{eq:estep}.  Even for the
current case where there are just two groups, this sum has $2^n$ terms and
would take prohibitively long to perform numerically for any but the
smallest of networks.  The most common way around this problem is to make
an approximate estimate of the sum by Monte Carlo sampling, but in this
paper we employ an alternative technique proposed by
Decelle~\etal~\cite{DKMZ11a,DKMZ11b}, which uses belief propagation.  This
technique is of interest both because it is significantly faster than Monte
Carlo and also because it lends itself to further analysis, as discussed in
Section~\ref{sec:detectability}.

Belief propagation~\cite{Pearl88}, a generalization of the Bethe--Peierls
iterative method for the solution of mean-field
models~\cite{Bethe35,Peierls36}, is a message-passing technique for finding
probability distributions on networks, which we can use in this case to
find the distribution~$q(g)$ of Eq.~\eqref{eq:estep}.  We define a
``message''~$\eta_r^{i\to j}$, which is equal to the probability that
vertex~$i$ belongs to group~$r$ if vertex~$j$ is removed from the network.
The removal of~$j$ allows one to derive a set of self-consistent of
equations that must be satisfied by these messages~\cite{DKMZ11a}.  The
equations are particularly simple for the case of a sparse network where
$p_{rs}$ is small so that terms of order~$p_{rs}$ can be ignored by
comparison with terms of order~1, which appears to describe most real-world
networks.  For this case, the equations are
\begin{equation}
\eta_r^{i\to j} = {\gamma_r\over Z_{i\to j}}
   \prod_k \biggl[ 1 - \sum_s q_s^k p_{rs} \biggr]
   \prod_{\substack{k(\ne j)\\ A_{ik}=1}}\sum_s \eta_s^{k\to i} p_{rs},
\label{eq:bp3}
\end{equation}
where~$Z_{i\to j}$ is a normalizing constant whose value is chosen to
ensure that $\sum_r \eta_r^{i\to j} = 1$ thus:
\begin{equation}
Z_{i\to j} = \sum_r \gamma_r \prod_k \biggl[ 1 - \sum_s q_s^k p_{rs} \biggr]
   \prod_{\substack{k(\ne j)\\ A_{ik}=1}}\sum_s \eta_s^{k\to i} p_{rs}.
\label{eq:z}
\end{equation}

Equation~\eqref{eq:bp3} is typically solved numerically, by starting from a
random initial condition and iterating to convergence.  In addition to
calculating new values for the messages~$\eta_r^{i\to j}$ on each step of
this iteration we also need to calculate new values for the one-vertex
marginal probabilities~$q_r^i$, which satisfy
\begin{equation}
q_r^i = {\gamma_r\over Z_i} \prod_k \biggl[ 1 - \sum_s q_s^k p_{rs} \biggr]
         \prod_{\substack{k\\A_{ik}=1}} \sum_s
         \eta_s^{k\to i} p_{rs},
\label{eq:onepoint3}
\end{equation}
with $Z_i$ being another normalization constant:
\begin{equation}
Z_i = \sum_r \gamma_r \prod_k \biggl[ 1 - \sum_s q_s^k p_{rs} \biggr]
   \prod_{\substack{k\\ A_{ik}=1}}\sum_s \eta_s^{k\to i} p_{rs}.
\label{eq:zi}
\end{equation}

Equation~\eqref{eq:bp3} is strictly true only on networks that are trees or
are \defn{locally tree-like}, meaning that in the limit of large network
size the neighborhood of any vertex looks like a tree out to arbitrarily
large distances.  The stochastic block model itself generates networks that
are locally tree-like, but many real-world networks are not, meaning that
the belief-propagation method is only approximate in those cases.  In
practice, however, it appears to give good results, comparable in quality
with those from Monte Carlo sampling (which is also an approximate method).

Once the belief propagation equations have converged, we can use the
results to evaluate Eq.~\eqref{eq:prs}.  This requires values of the
two-vertex marginals, which are given by Bayes theorem to be
\begin{align}
q_{rs}^{ij} &= P(g_i=r,g_j=s|A_{ij}=1) \nonumber\\
            &= {P(g_i=r,g_j=s)\over P(A_{ij}=1)} P(A_{ij}=1|g_i=r,g_j=s),
\end{align}
where all elements of the adjacency matrix other than~$A_{ij}$ are assumed
given in each probability.  In terms of our other variables we have
\begin{align}
P(g_i=r,g_j=s) &= \eta_r^{i\to j} \eta_s^{j\to i}, \nonumber\\
P(A_{ij}=1|g_i=r,g_j=s) &= p_{rs},
\end{align}
and the normalization~$P(A_{ij})$ is fixed by the requirement that
$q_{rs}^{ij}$ sum to unity.  So
\begin{equation}
q_{rs}^{ij} = {\eta_r^{i\to j} \eta_s^{j\to i} p_{rs}\over
               \sum_{rs} \eta_r^{i\to j} \eta_s^{j\to i} p_{rs}}.
\label{eq:marginals}
\end{equation}
Substituting the values of $q_r^i$ and $q_{rs}^{ij}$ into
Eqs.~\eqref{eq:gamma} and~\eqref{eq:prs} then completes the EM algorithm.

Note that there are now two entirely separate iterative sections of our
calculation: the EM algorithm, which consists of the iteration of
Eqs.~\eqref{eq:estep}, \eqref{eq:gamma}, and~\eqref{eq:prs}, and the belief
propagation algorithm, which consists of the iteration of
Eq.~\eqref{eq:bp3}.

Using the belief propagation algorithm is far faster than
calculating~$q(g)$ directly from Eq.~\eqref{eq:estep}.
Equations~\eqref{eq:bp3} to~\eqref{eq:zi} require the evaluation of only
$\Ord(m+n)$ terms for a network with $n$ vertices and $m$ edges, meaning an
iteration takes linear time in the common case of a sparse network with
$m\propto n$.  There is still the issue of how many iterations are needed
for convergence, for which there are no firm results at present, but
heuristic arguments suggest that $\Ord(\log n)$ iterations will be needed
on a typical network.

The complete algorithm for detecting core--periphery structure in networks
consists of the following steps:
\begin{enumerate}
\item Make an initial random guess at the values of the
  parameters~$p,\gamma$.
\item From a random initial condition, iterate to convergence the belief
  propagation equations~\eqref{eq:bp3} for vertex pairs connected by an
  edge and the one-vertex marginal probabilities, Eq.~\eqref{eq:onepoint3}.
\item Use the converged values to calculate the two-vertex marginal
  probabilities, Eq.~\eqref{eq:marginals}.
\item Use the one- and two-vertex probabilities to calculate an improved
  estimate of the parameters from Eqs.~\eqref{eq:gamma} and~\eqref{eq:prs}.
\item Repeat from step~2 until the parameters converge.
\item Assign each vertex to either the core or the periphery, whichever has
  the higher probability~$q_r^i$.
\end{enumerate}

\section{Detectability}
\label{sec:detectability}
One of the most intriguing aspects of the community detection problem is
the \defn{detectability threshold}~\cite{RL08,DKMZ11a,HRN12}.  When a
network contains strong community structure---when there is a clear
difference in density between the in-group and out-group connections---then
that structure is easy to detect and a wide range of algorithms will do a
good job.  When structure becomes sufficiently weak, however, at least in
simple models of the problem such as the stochastic block model, it becomes
undetectable.  In this weak-structure regime it is rigorously provable that
no algorithm can identify community memberships with success any better
than a random coin toss~\cite{MNS12,MNS13}.  Given the strong connection
between community detection and the core--periphery detection problem
studied here, it is natural to ask whether there is a similar threshold for
the core--periphery problem.  Is there a point at which core--periphery
structure becomes so weak as to be undetectable by our method or any other?

At the most naive level, the answer to this question is no.  The
core--periphery problem differs from the community detection problem in
that the vertices in the core have higher degree on average than those in
the periphery and hence one can use the degrees to identify the core and
periphery vertices with an average success rate better than a coin toss.

\begin{figure}
\begin{center}
\includegraphics[width=\columnwidth]{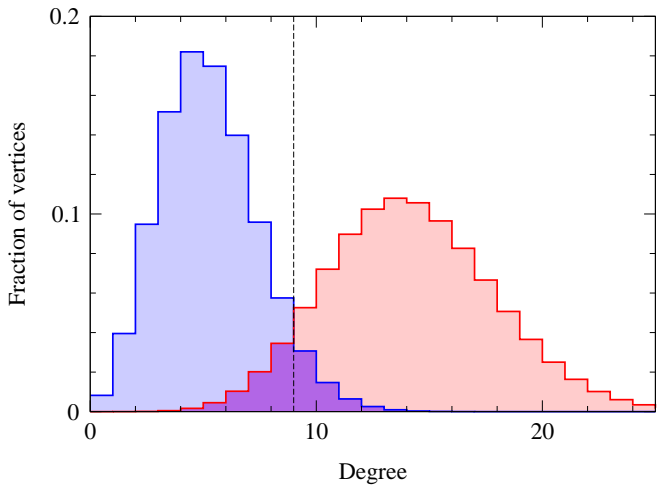}
\end{center}
\caption{In the stochastic block model both core (red) and periphery (blue)
  vertices have Poisson degree distributions, but the mean degree is higher
  in the core than in the periphery, so the overall degree distribution of
  the network is a sum of two overlapping Poisson distributions as shown
  here.  A simple division of vertices by degree (vertical dashed line)
  classifies most vertices into the correct groups, red in the core and
  blue in the periphery.  Only those in the overlap (shown in purple) are
  classified incorrectly.}
\label{fig:poisson}
\end{figure}

Consider in particular the common case of a stochastic block model where
\begin{equation}
p_{rs} = {c_{rs}\over n}
\end{equation}
for some constants~$c_{rs}$.  This is the case for which the detectability
threshold mentioned above is observed.  Then the average degrees in the
core and periphery are, respectively,
\begin{equation}
\bar{d}_1 = \gamma_1 c_{11} + \gamma_2 c_{12}, \qquad
\bar{d}_2 = \gamma_1 c_{12} + \gamma_2 c_{22},
\label{eq:bard}
\end{equation}
and the difference is $\bar{d}_1 - \bar{d}_2 = \gamma_1(c_{11}-c_{12}) +
\gamma_2(c_{12}-c_{22})$.  Since, by hypothesis, $c_{11}>c_{12}>c_{22}$,
this quantity is always positive and $\bar{d}_1>\bar{d}_2$.  Because the
edges in the network are independent, the actual degrees have a Poisson
distribution about the mean in the limit of large~$n$, and hence the degree
distribution consists of two overlapping Poisson distributions, as sketched
in Fig.~\ref{fig:poisson}.  By simply dividing the vertices according to
their observed degrees, therefore, we can (on average) classify them as
core or periphery with success better than chance.  (This assumes we know
the sizes of the two groups, which we usually don't, but this problem can
be solved---see Section~\ref{sec:degreeonly}.)

So rather than asking whether our ability to detect structure fails
completely in the weak-structure limit, we should instead ask whether we
can do any better than simply dividing vertices according to degree.  The
answer to this question is both yes and no.  As we now show, in the limit
of weak structure no algorithm can do better than one that looks at degrees
only, but for stronger structure we can do better in most cases.

To demonstrate these results, we take a standard approach from statistics
and ask whether our detection algorithm based on the stochastic block model
can detect core--periphery structure in networks that are themselves
generated using the stochastic block model.  This is a so-called
consistency test and, in addition to providing a well-controlled test of
our algorithm, it has one very important advantage.  It is known that on
average the best way to detect the structure in a data set generated by a
model is to perform a maximum-likelihood fit to that same model, exactly as
our algorithm does.  No other algorithm will return better performance on
this test, on average, than the maximum likelihood method.

Bearing this in mind, consider applying the algorithm of this paper to a
network generated using the stochastic block model with two equally sized
groups ($\gamma_1=\gamma_2=\frac12$) and weak core--periphery structure of
the form
\begin{equation}
c_{11} = c + \alpha_1\delta, \quad c_{12} = c, \quad
c_{22} = c - \alpha_2\delta,
\label{eq:delta}
\end{equation}
where $\alpha_1$, $\alpha_2$, and~$c$ are $\Ord(1)$ positive constants and
$\delta$~is a small quantity.  In the limit as $\delta\to0$ the
core--periphery structure vanishes and the network becomes a uniform random
graph of average degree~$c$.  For small values of~$\delta$ the structure is
weak and it is this regime that we're interested to probe.

To make the problem as simple as possible, suppose that we allow our
algorithm to use the exact values of the parameters~$\gamma_r$
and~$p_{rs}$, meaning that we need only perform the belief propagation part
of the calculation to derive an answer.  There is no need to perform the EM
algorithm iteration as well, since this is only needed to determine the
parameters.  This is a somewhat unrealistic situation---in practical cases
we do not normally know the values of the parameters.  However, if, as we
will show, the algorithm performs poorly in this situation then it will
surely perform no better if we give it \emph{less} information---if we do
not know the values of the parameters.  Thus this choice gives us a
best-case estimate of the performance of the algorithm.

To gain a theoretical understanding of how the belief propagation process
works, we consider the \defn{odds ratio} $q_1^i/q_2^i$ between the
probabilities that a vertex belongs to the core and the periphery.  Making
use of Eq.~\eqref{eq:onepoint3}, expanding the first product to leading
order in $p_{rs}=c_{rs}/n$, and dividing top and bottom in the second
product by a factor of~$n$, this quantity is given by
\begin{equation}
{q_1^i\over q_2^i} = {\gamma_1\over\gamma_2}\,\e^{\bar{d}_2-\bar{d}_1}\!
  \prod_{\substack{k\\ A_{ik}=1}}
  {\eta_1^{i\to j}c_{11} + \eta_2^{i\to j}c_{12}\over
  \eta_1^{i\to j} c_{12} +\eta_1^{i\to j} c_{22}},
\label{eq:phi}
\end{equation}
where $\bar{d}_1$ and $\bar{d}_2$ are defined as in Eq.~\eqref{eq:bard} and
we have made use of Eq.~\eqref{eq:gamma}.  Note how the
normalization~$Z_{i\to j}$ also cancels, making calculations simpler.

Now we substitute for $c_{rs}$ from Eq.~\eqref{eq:delta}, set
$\gamma_1=\gamma_2=\frac12$, and note that as $\delta\to0$ the
probabilities of any vertex being in one group or the other become equal,
so that
\begin{equation}
{\eta_1^{i\to j}\over\eta_2^{i\to j}} = 1 + \beta_{i\to j} \delta
\end{equation}
to leading order for some constant~$\beta_{i\to j}$.  Keeping terms to
first order in~$\delta$, we then find that
\begin{equation}
{q_1^i\over q_2^i} = 1 + \tfrac12(\alpha_1+\alpha_2){k_i-c\over c} \delta,
\end{equation}
where $k_i$ is the degree of vertex~$i$ as previously.

Note that $\beta_{i\to j}$ has dropped out of this expression, meaning that
when $\delta$ is small and the structure is weak the probabilities depend
only on the degree~$k_i$ of the vertex and not on any other properties of
the network structure.  More specifically, vertex~$i$ has a higher
probability of belonging to group~1, i.e.,~the core, whenever its
degree~$k_i$ is greater than the average degree~$c$ in the network as a
whole.  When its degree is below average the vertex has a higher
probability of belonging to the periphery.  Thus a simple division based on
probabilities is precisely equivalent to dividing based on degree.
Moreover, since, as we have said, no other algorithm can do better at
distinguishing the structure, it immediately follows that there is nothing
better one can do in the weak-structure limit than divide the vertices
based on degree.

Paradoxically, the same is also true in the limit of strong structure.  If
the core--periphery structure is strong, meaning that there is a big
difference between connection probabilities for core and periphery
vertices, then the two Poisson distributions of Fig.~\ref{fig:poisson} will
be far apart, with very little overlap, and vertices can be accurately
classified by degree alone.  The means of the two distributions are
$\mu_1=\frac12(c_{11}+c_{12})$ and $\mu_2=\frac12(c_{12}+c_{22})$ and,
since the width of a Poisson distribution scales as the square root of its
mean, we will have easily distinguishable peaks provided
$\mu_1-\mu_2\gg\sqrt{\smash[b]{(\mu_1+\mu_2)/2}}$, or
\begin{equation}
c_{11}-c_{22} \gg 2\sqrt{c},
\end{equation}
where $c=\frac12(\mu_1+\mu_2)$ is the average degree of the network as a
whole.

In fact, even between the limits of strong and weak structure there are
some networks for which a simple division by degrees is optimal.  Consider
the two-parameter family of models defined by
\begin{equation}
c_{11} = \theta r, \quad c_{12} = \theta, \quad c_{22} = {\theta\over r},
\label{eq:plane}
\end{equation}
for any choice of~$\gamma_r$, where $\theta>0$ and $r>1$.  Substituting
this choice into Eq.~\eqref{eq:phi} gives
\begin{equation}
{q_1^i\over q_2^i} = {\gamma_1\over\gamma_2}\,\e^{\bar{d}_2-\bar{d}_1}
                     r^{k_i},
\label{eq:odds}
\end{equation}
so again the results depend only on the vertex degrees.

So are there any cases where we can do better than the algorithm that looks
at degrees only?  The answer is yes: for structure of intermediate
strength, neither exceptionally weak nor exceptionally strong, and away
from the plane in parameter space defined by Eq.~\eqref{eq:plane}, the
messages are not simple functions of degree but depend in general on the
details of the network structure.  Since, once again, the belief
propagation algorithm is optimal, it follows that any algorithm that gives
a result different from the belief propagation algorithm must give an
inferior one, including an algorithm that looks at degrees only.  Hence in
this regime one can do better than simply looking at vertex degrees.
Moreover, this regime contains most cases of real-world interest.  After
all, core--periphery structure so weak as to be barely detectable is
presumably not of great interest, and real-world networks rarely have
strongly bimodal degree distributions of the kind considered above that
make degree-based algorithms work well in the strong-structure limit.

There is also, we note, no evidence in this case of a detectability
threshold or similar sharp discontinuity in the behavior of the algorithm.
Everywhere in the parameter space the algorithm can identify core and
periphery with performance better than chance.

\subsection{Degree-based algorithm}
\label{sec:degreeonly}
We are now also in a position to answer a question raised parenthetically
in Section~\ref{sec:bp}.  If we choose to classify vertices based on degree
alone, what size groups should we use?  We can answer this question by
noting that Eq.~\eqref{eq:plane} defines the subset of stochastic block
models for which degree alone governs classification.  As we have seen,
fitting to this model is equivalent to dividing according to degree, but
performing such a fit rather than just looking at degrees has the added
advantage that it gives us the values of the parameters~$\gamma_r$, which
in turn give us the expected sizes $n_r = n\gamma_r$ of the groups.  We can
perform the fit exactly as we did for the full stochastic block model in
Section~\ref{sec:em}.  Substituting Eq.~\eqref{eq:plane} into the
right-hand side of~\eqref{eq:ineq}, differentiating, and neglecting terms
of order~$1/n$ by comparison with those of order~1, we find the optimal
values of the parameters to be
\begin{equation}
\gamma_r = {1\over n} \sum_i q_r^i, \qquad
r = {\kappa_1\over\kappa_2}, \qquad
\theta = {\kappa_1\kappa_2\over c},
\label{eq:degree1}
\end{equation}
where $c$ is the average degree of the network as previously and $\kappa_r$
is the expected degree in group~$r$:
\begin{equation}
\kappa_r = {\sum_i k_i q_r^i\over\sum_i q_r^i}.
\end{equation}
The one-vertex probabilities $q_r^i$ are given by Eq.~\eqref{eq:odds} to be
\begin{equation}
q_1^i = {\gamma_1\e^{-\bar{d}_1} r^{k_i}\over\gamma_2\e^{-\bar{d}_2}
         + \gamma_1\e^{-\bar{d}_1} r^{k_i}}, \qquad
q_2^i = 1 - q_1^i.
\label{eq:degree2}
\end{equation}

Hence for this model, no belief propagation is necessary.  One can simply
iterate Eqs.~\eqref{eq:degree1} and~\eqref{eq:degree2} to convergence to
determine the group memberships.  (Note that in fact the parameter~$\theta$
is never needed in the iteration---it is sufficient to calculate only
$\gamma_1$, $\gamma_2$, and~$r$ from Eq.~\eqref{eq:degree1}.)

\section{Applications and performance}
\label{sec:examples}
We have tested the proposed method on both computer-generated and
real-world example networks.

\subsection{Computer-generated test networks}
Computer-generated networks provide a controlled test of the algorithm's
ability to detect known structure.  For these tests we make use of the
stochastic block model itself to generate the test networks.  We
parametrize the mixing matrix of the model as
\begin{equation}
\begin{pmatrix}
c_{11} & c_{12} \\
c_{21} & c_{22}
\end{pmatrix}
= \theta_1 \vec{u}_1\vec{u}_1^T + \theta_2 \vec{u}_2\vec{u}_2^T,
\label{eq:rtheta}
\end{equation}
where $\vec{u}_1 = (\sqrt{r},1/\sqrt{r})$ and $\vec{u}_2 =
(1/\sqrt{r},-\sqrt{r})$.  With this parametrization, setting $\theta_2=0$
recovers the $(\theta,r)$-model of Section~\ref{sec:detectability}, for
which, as we showed there, no algorithm does any better than a naive
division according to vertex degree only.  The parameter~$\theta_2$
measures how far away we are from that model in the perpendicular direction
defined by~$\vec{u}_2$, and we might guess that when we are further
away---i.e.,~for values of~$\theta_2$ further from zero---we would see a
greater difference between the belief propagation algorithm and the naive
one.

\begin{figure}
\begin{center}
\includegraphics[width=\columnwidth]{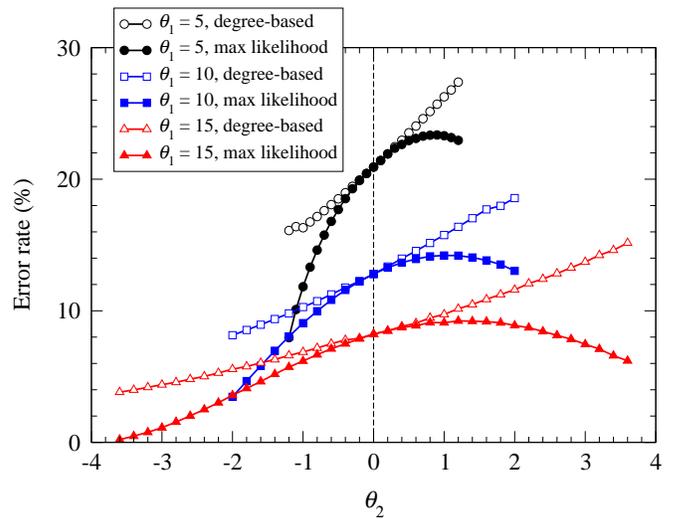}
\end{center}
\caption{The fraction of nodes classified incorrectly in tests on
  stochastic block model networks parametrized according to
  Eq.~\eqref{eq:rtheta}, as a function of $\theta_2$ for fixed~$r=2$ and
  three different values of~$\theta_1$ as indicated.  Solid points
  represent results for the maximum likelihood method described in this
  paper.  Open points are the results of a simple division according to
  vertex degree.  Each point is an average over 10 networks of a million
  nodes.  Statistical errors are smaller than the data points.  The
  parameter ranges are different for different curves because they are
  constrained by the requirement that edge probabilities be nonnegative and
  that $c_{11}>c_{12}>c_{22}$, which means that $\theta_2$ must satisfy
  $-\theta_1/r<\theta_2<\theta_1(1-1/r)/(r+1)$.}
\label{fig:errorrate}
\end{figure}

Figure~\ref{fig:errorrate} shows this indeed to be the case.  The figure
shows, for three different choices of~$\theta_1$, the error rate of the
algorithm (i.e.,~the fraction of incorrectly identified vertices) as a
function of~$\theta_2$ for networks of $n=1\,000\,000$ nodes, divided into
equally sized core and periphery.  Also shown on the plot is the
performance of the algorithm that simply divides the vertices into two
equally sized groups according to degree.  As we can see, when $\theta_2=0$
(marked by the vertical dashed line) the results for the two approaches
coincide as we expect.  But as $\theta_2$ moves away from zero there is a
visible difference between the two, with the error rate of the naive
algorithm being worse than that of belief propagation by a factor of ten or
more in some cases.

It is fair to say, however, that the error rates of the two algorithms are
comparable in some cases and the naive algorithm does moderately well under
the right conditions, with error rates of around 10 or 20 percent for many
choices of parameter values.  There are a couple of possible morals one can
derive from this observation.  On the one hand, if one is not greatly
concerned with accuracy and just wants a quick-and-dirty division into core
and periphery, then dividing vertices by degree may be a viable strategy.
The belief propagation method usually does better, but it is also more work
to program and requires more CPU time to execute.  For some applications we
may feel that the additional effort is not worth the payoff.  Moreover,
since the belief propagation method is optimal in the sense discussed
earlier, we know that, at least for the definition of core--periphery
structure used here, no other algorithm will out-perform it, so the loss of
accuracy seen in Fig.~\ref{fig:errorrate} is the largest such loss we will
ever incur when using the degree-based algorithm.  In other words, this is
as bad as it gets, and it's not that bad.

On the other hand, as we have said, one does not in most cases know the
sizes of the groups into which the network is to be divided, in which case
one must use the EM algorithm even for a degree-based division.  The
computations involved, which are described in Section~\ref{sec:degreeonly},
are less arduous than those for the full belief propagation algorithm but
significantly more complex than a simple division by degree only, and this
eliminates some of the advantages of the degree-based approach.

Furthermore, while the number of nodes on which our method and the
degree-based algorithm differ is sometimes quite small, it may be these
very nodes that are of greatest interest.  It's true that it is typically
the higher-degree nodes that fall in the core and the lower-degree ones
that fall in the periphery.  But when the two algorithms differ in their
predictions it is precisely because some of the low-degree nodes correctly
belong in the core or some of the high-degree ones in the periphery, which
could lead us to ask what is special about these nodes.  Who are the people
in a social network, for example, who fall in the core even though they
don't have many connections?  Who are the well-connected people who fall in
the periphery?  These people may be of particular interest to us, but they
can only be identified by using the full maximum likelihood algorithm.  The
degree-based algorithm will, by definition, never find these anomalous
nodes.

\subsection{Real-world examples}
\label{sec:example} 
\begin{figure}
\begin{center}
\includegraphics[width=\columnwidth]{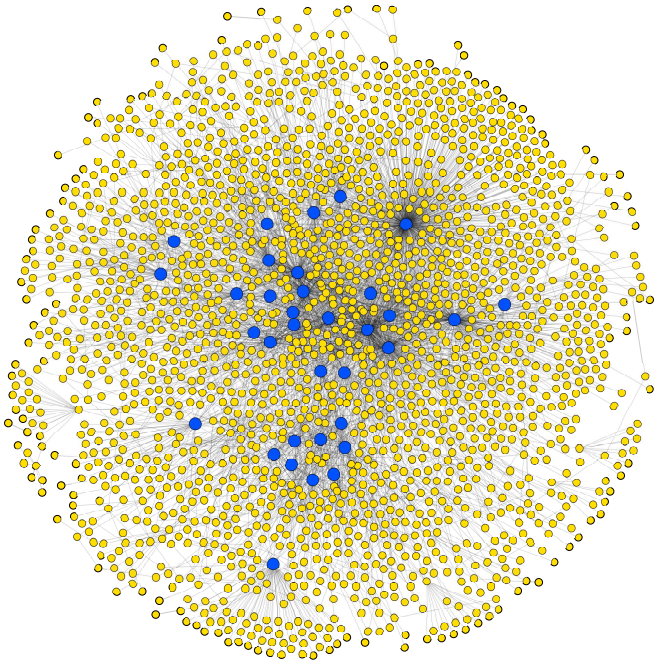}
\end{center}
\caption{Core--periphery division of a 1470-node representation of the
  Internet at the level of autonomous systems~\cite{PVV01}.  Nodes placed
  in the core by our analysis are drawn larger and in blue; nodes in the
  periphery are smaller and in yellow.  The network was constructed from
  data from the Oregon Routeviews Project and represents an older snapshot,
  chosen for the network's relatively small size.  Our methods can easily
  be applied to larger networks, but smaller size makes the visualization
  of the results clearer.}
\label{fig:internet}
\end{figure}

Figure~\ref{fig:internet} shows an application of our method to a
real-world network, the Internet, represented at the level of autonomous
systems.  This network is expected to have clear core--periphery structure:
its general structure consists of a large number of leaves or edge
nodes---typically client autonomous systems corresponding to end users like
ISPs, corporations, or educational institutions---plus a smaller number of
well-connected backbone nodes~\cite{PV04,Holme05b}.  This structure is
reflected in the decomposition discovered by our analysis, indicated by the
blue (core) and yellow (periphery) nodes in the figure.  The bulk of the
nodes are placed in the periphery, while a small fraction of central hubs
are placed in the core.  Note, however, that, as discussed earlier, the
algorithm does not simply divide the nodes according to degree.  There are a
significant number of high-degree nodes that are placed by the algorithm in
the periphery because of their position on the fringes of the network, even
though their degree might naively suggest that they be placed in the core.

\begin{figure}
\begin{center}
\includegraphics[width=\columnwidth]{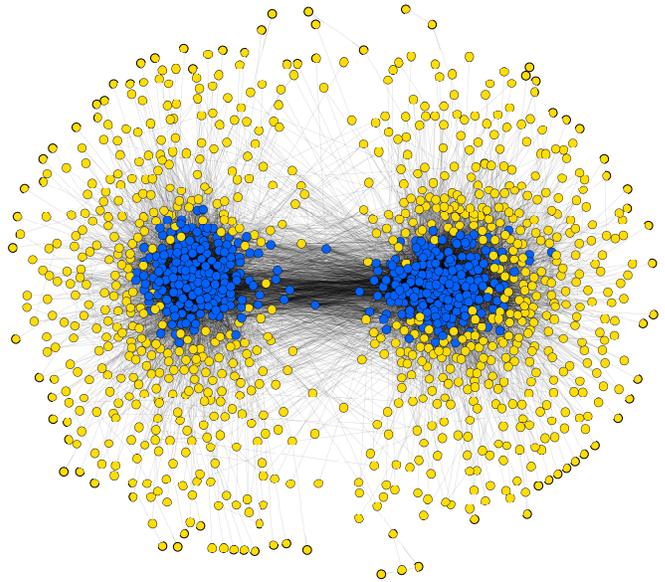}
\end{center}
\caption{Core--periphery division of a network of hyperlinks between
  political blogs taken from~\cite{AG05}.  The network naturally separates
  into conservative and liberal communities, clearly visible as the two
  clusters in this picture.  Within each group our algorithm finds a
  separate core and periphery indicated by the blue and yellow nodes
  respectively.}
\label{fig:polblogs}
\end{figure}

Figure~\ref{fig:polblogs} shows a contrasting example.  The network in this
figure, drawn from a 2005 study by Adamic and Glance~\cite{AG05} is a web
network, representing a set of 1225 weblogs, personal commentary web sites,
devoted in this case to commentary on US politics.  Edges represent
hyperlinks between blogs, which we treat as undirected for the purposes of
our analysis.  This network has been studied previously as an example of
community structure, since it displays a marked division into groups of
conservative and liberal blogs.  The figure is drawn so as to make these
groups clear to the eye---they correspond roughly to the left and right
halves of the picture---and the core--periphery division is indicated once
more by the blue (core) and yellow (periphery) nodes.

As the figure shows, the analysis finds a clear separation between core and
periphery, and moreover finds a separate core in each of the two
communities.  In effect, the conservative blogs are divided into a
conservative core and periphery, and similarly for the liberal ones.  A
direct examination of the list of core nodes in each community finds them
to contain, as we might expect, many prominent blogs on either side of the
aisle, such as the National Review and Red State on the conservative side
and Daily Kos and Talking Points Memo on the liberal side.

\section{Conclusion}
We have examined core--periphery structure in undirected networks,
proposing a first-principles algorithm for identifying such structure by
fitting a stochastic block model to observed network data using a maximum
likelihood method.  The maximization is implemented using a combination of
an expectation--maximization algorithm and belief propagation.  The
algorithm gives good results on test networks and is efficient enough to
scale to networks of a million nodes or more.  By a linearization of the
belief propagation equations we are also able to show the method to be
immune from the detectability threshold seen in the application of similar
methods to community detection.  In the community detection case the
algorithm (and indeed all algorithms) fail when community structure in the
network is too weak, but there is no such failure for the core--periphery
case.  Core--periphery structure is always detectable, no matter how weak
it is.

\begin{acknowledgments}
  The authors thank Cris Moore for useful conversations and Petter Holme
  for providing the network data for Fig.~\ref{fig:internet}.  This work
  was funded in part by the National Science Foundation under grants
  DMS--1107796 and DMS--1407207 and by the Air Force Office of Scientific
  Research (AFOSR) and the Defense Advanced Research Projects Agency
  (DARPA) under grant FA9550--12--1--0432.
\end{acknowledgments}

\bibliographystyle{numeric}
\bibliography{journals,references}

\end{document}